# Deliberative Democracy, Perspective from Indo-Pacific Blogosphere: A Survey

ABIOLA AKINNUBI*
NITIN AGARWAL

Collaboratorium for Social Media and Online Behavioral Studies, USA

## Abstract

Deliberation and communication within the national space have had numerous implications on how citizens online and offline perceive government. It has also impacted the relationship between opposition and incumbent governments in the Indo-Pacific region. Authoritarian regimes have historically had control over the dissemination of information, thereby controlling power and limiting challenges from citizens who are not comfortable with the status quo. Social media and blogs have allowed citizens of these countries to find a way to communicate, and the exchange of information continues to rise. The quest by both authoritarian and democratic regimes to control or influence the discussion in the public sphere has given rise to concepts like cybertroopers, congressional bloggers, and commentator bloggers, among others. Cybertroopers have become the de facto online soldiers of authoritarian regimes who must embrace democracy. While commentator and congressional bloggers have acted with different strategies, commentator bloggers educate online citizens with knowledgeable information to influence the citizens. Congressional bloggers are political officeholders who use blogging to communicate their positions on ongoing national issues. Therefore, this work has explored various concepts synonymous with the Indo-Pacific public sphere and how it shapes elections and democracy.

## 1. Introduction

The concept of deliberative democracy has historically been why democratic society like the United States wants every nation to follow in its footsteps. This is because citizens will have the power and medium to discuss salient issues about their interests and challenge constituted authorities in their country. While some countries in the Indo-Pacific region have democratically elected governments, many Indo-Pacific nations have hidden behind democracy and elections to run autocratic governments. Thereby suppressing opposing views and polarizing the blogosphere and social media, where citizens have gathered in numbers to discuss issues affecting their countries. Citizens have also begun channeling their emotions through blogs and social media. Since various paid media like newspapers and public television stations have now become the running town-crier arm of the government at the center. The press, which the government regulates, has bowed to suppression due to licensing and government patronage by not promoting reality on the ground. The opposition has resorted to blogs and social media to voice their opinions than using conventional media platforms. Many would argue that the opposition should have their news media. Still, with government licensing acting as a barrier, the opposition interested in power has fewer options than to run to social media and blogs where the communities are to push their narratives.

However, participatory democracy comes with the difficult challenge of ensuring that polarization does not dissuade the conversation from critical economic and socio-political issues facing any country in the Indo-Pacific region. Many traditional autocratic regimes who have since transitioned to democratic governments have recruited agents and actors to infiltrate intelligent conversations online through blogging or other platforms like Twitter, Facebook, and Reddit. These intruders in the public sphere are popularly described as **cybertroopers**. They are in the conversation to disrupt or push narratives that debate the actual discussion for the benefit of their sponsors. Many elected officials have engaged in public debates themselves, popularly described as **congressional bloggers**. It allows them to enter the fray of conversation to take direct feedback and gauge the general mood when everyone is in the same room via comments and reading blog posts from the public officeholders.

Also, the public sphere has **influential bloggers** most trusted when commenting on political and social issues. **Influential or commentator bloggers** can provide a detailed analysis of issues, commonly referred to as **influential or commentator bloggers**. But, since the Indo-Pacific region consists of different countries, they have a complex democratic practice unique to each. This survey uses Google scholar, ResearchGate, Web of Science, and Academia to follow the various published resources due to their vast data availability for multiple studies and easy access to journals. However, literature has also shown that they share certain similarities. This work studies and extrapolates the contemplative nature of the Indo-Pacific public sphere. More specifically, this study analyzes the type of bloggers, how the political ruling class can enter the public sphere by diffusing conversation, and how the opposition uses blogs and the public sphere in socio-political debates in the Indo-Pacific region.

## 2. Methodology

This survey uses Google scholar, ResearchGate, Web of Science, and Academia to follow the various published resources due to their vast data availability for multiple studies and easy access to journals with limited restrictions. We filtered the results to countries whose democracy and public sphere deliberation around governance have been keenly influenced by adopting blogs as a means of citizen participation and authority participation in democracy. We used the Boolean query in Figure 1 with phrases around Indo-Pacific Deliberative democracy, blogger types, and other key issues when searching for journal articles. Adopting this approach helps ensure that there is little noise in the search results from various sources from which we collected the data. We specifically look for journals and articles about the blogosphere with mentions and focus on any countries on the map shown in Figure 2 and Table 1 below.

Collecting data using the map in Figure 2 helps ensure that the quality of materials returned from the search is as close as possible to the Indo-Pacific region countries. Furthermore, we described using a high-level flow diagram in Figure 3 to show how we studied the collected data. Figure 3 also shows other tools used in the study for journals and article management, like Zotero, while we used vosviewer and Gephi for network analysis. Two hundred and twelve articles starting from the year 1999 to 2022 to better categorize the paper into appropriate themes, namely, democracy, religion, and politics, and also to categorize bloggers in the Indo-Pacific region according to their behavior which is: cybertroopers which are bloggers that help manipulate online opinion by countering or pushing narratives for their paid benefactors, commentator: these are bloggers actively online to educate in an online discussion due to their interest in the topic of discussion or wealth of knowledge in the domain. They are usually not sponsored or paid to do so, and congressional bloggers are mainly political officeholders who blog about their views on a national issue.

The categorization approach was used to ensure that the various papers are grouped by the key themes they discussed, as briefly described in a few sentences that precede this and the country they belong to. Although it was observed that there is a cross-reference between various themes, we were able to proceed with this categorization since they are all related and belong to either the same country or theme. We use vosviewer to analyze bibliography and author networks to correlate resources that various work studies have drawn motivation and see how this helps our work understand the transition of multiple discussions and work around the study region.

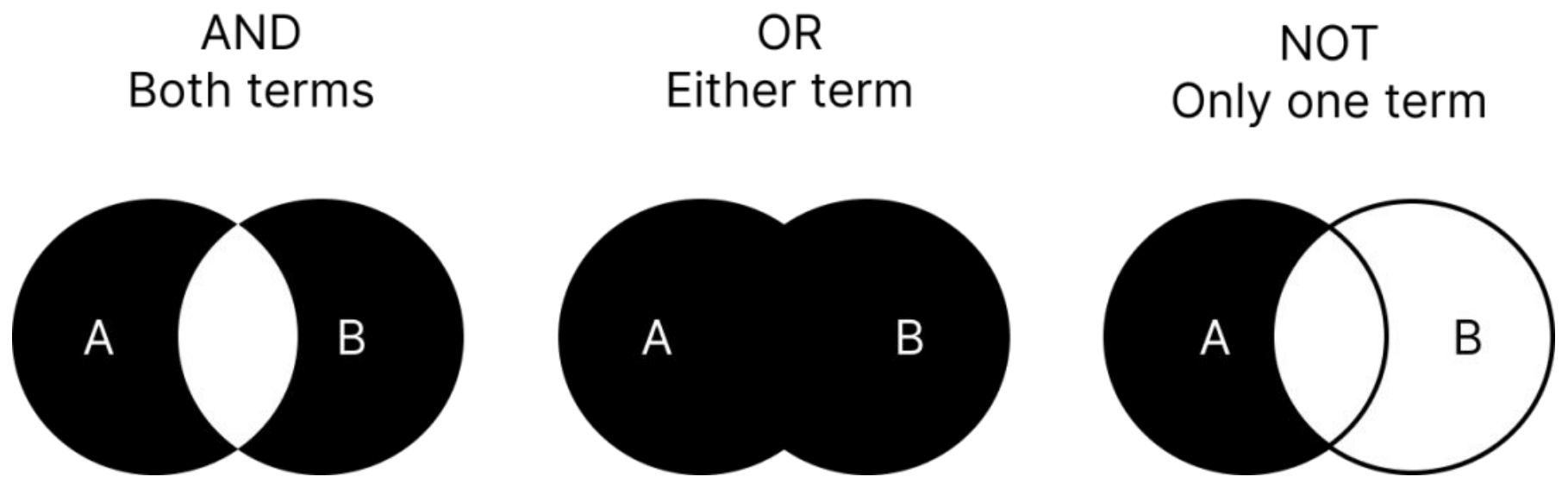

*Figure 1: Boolean logic for article search*

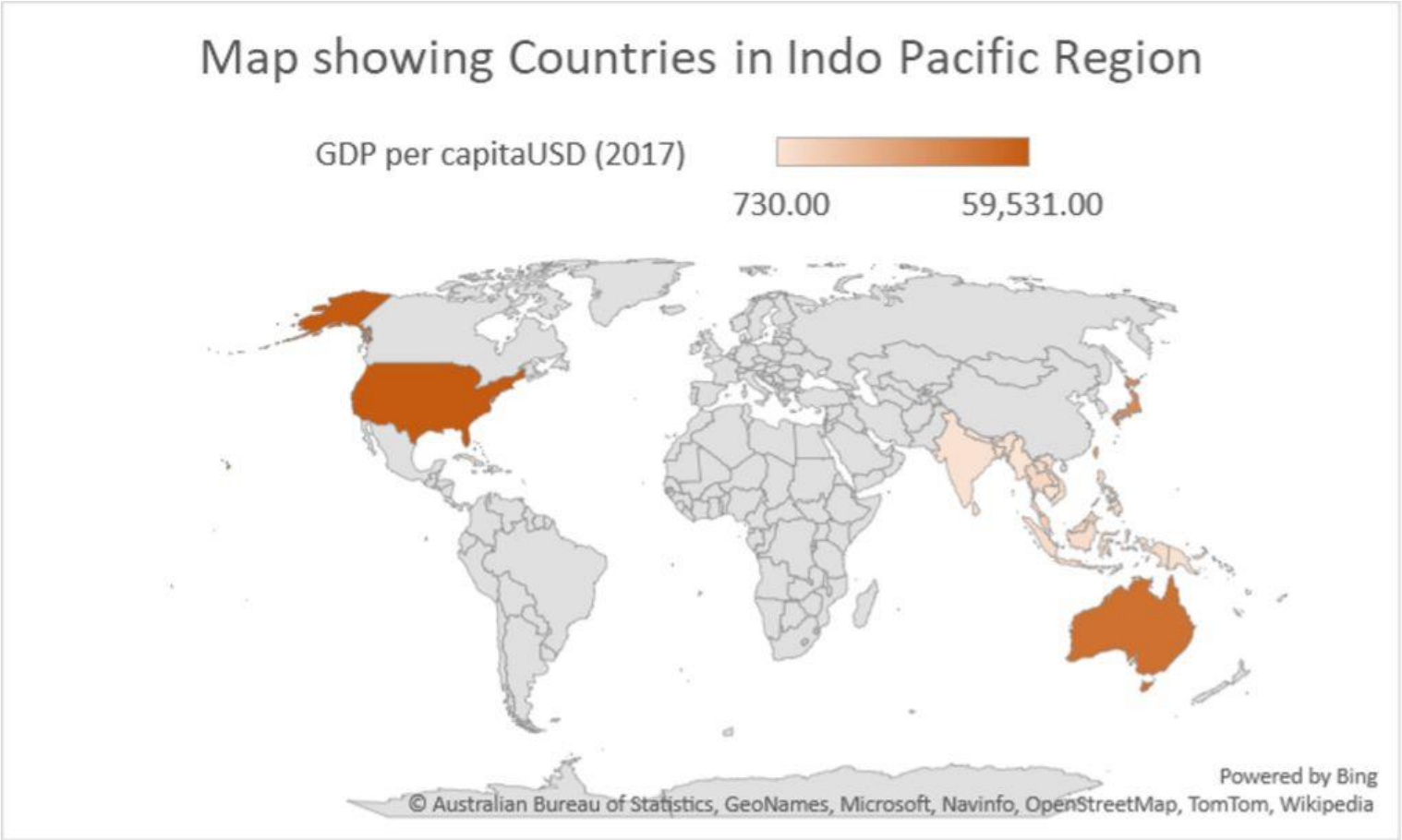

*Figure 2: Map showing per capital of highlighted Indo-Pacific countries*

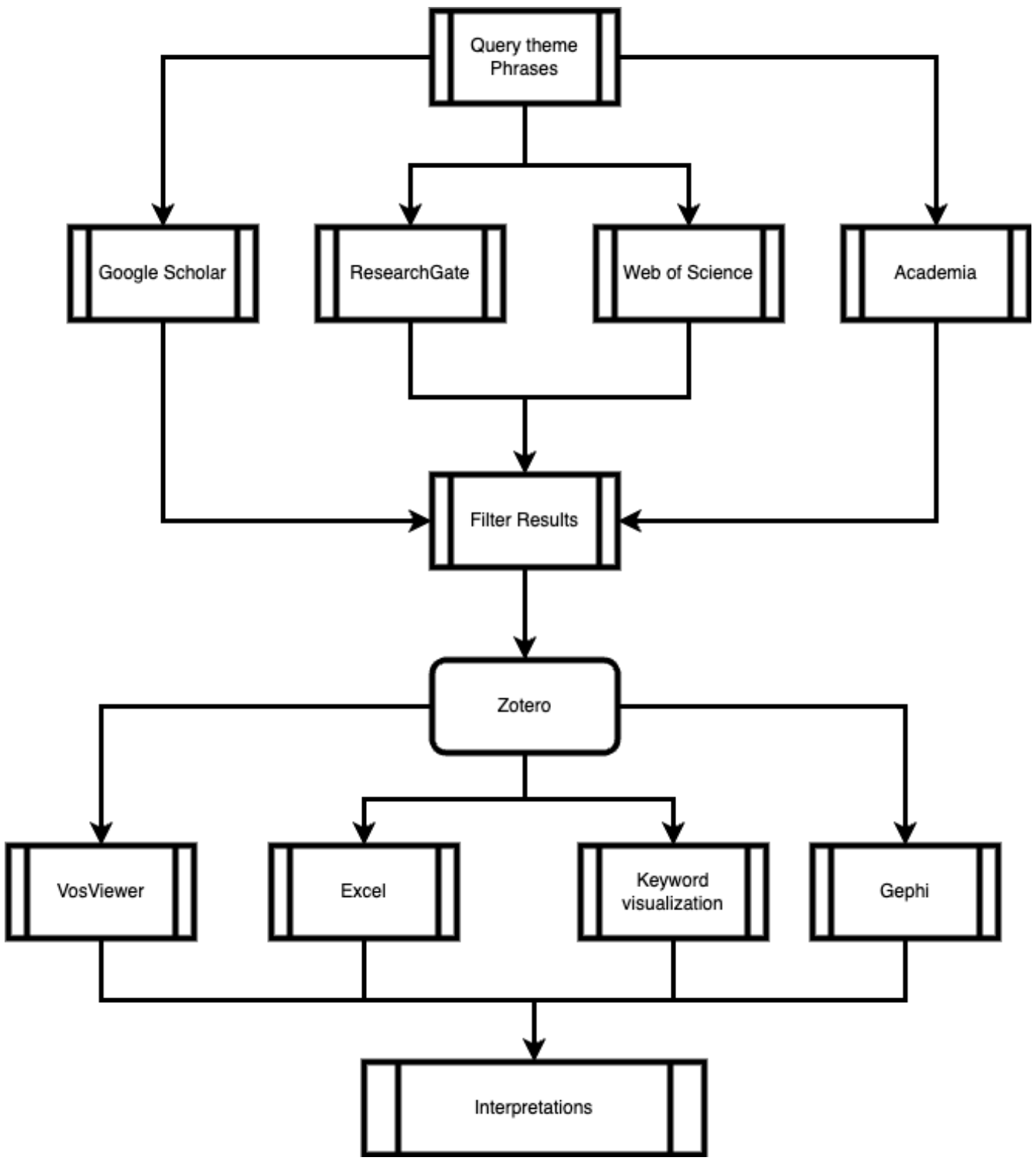

*Figure 3: Methodology and data collection used in this study*

*Table 1: Table showing Indo-Pacific counties and their pa-capital GDP along with other statistics adapted from https://ceoworld.biz/Indo-Pacific/ accessed Jun 8, 2*

| Country | Land Sq | Population | GDP (2017) | Per Capital USD |
|---|---|---|---|---|
| India | 3,287,263 | 1,324,171,354 | 2,690,000 | 1,939 |
| United States | 9,833,520 | 328,271,859 | 19,386,800 | 59,531 |
| Indonesia | 1,904,569 | 252,164,800 | 856,066 | 3,570 |
| Bangladesh | 147,570 | 156,594,962 | 205,715 | 1,359 |
| Japan | 377,944 | 126,434964 | 4,769,804 | 38,894 |
| Philippines | 343,448 | 107,242,00 | 289,686 | 2,951 |
| Vietnam | 331,210 | 88,069,000 | 187,848 | 2,186 |
| Thailand | 5,13,120 | 67,764,000 | 380,491 | 5,908 |
| Myanmar | 6,76,578 | 50,496,000 | 65,291 | 1,275 |
| Malaysia | 329,847 | 30,185,787 | 336,913 | 9,503 |
| Nepal | 147,181 | 26,494,504 | 62,384 | 730 |
| Australia | 7,692,024 | 23,731,000 | 1,482,539 | 49,928 |
| Taiwan | 36,191 | 23,119,772 | 505,452 | 31,900 |
| Sri Lanka | 65,610 | 20,277,597 | 233,637 | 3,835 |
| Cambodia | 181,035 | 15,205,539 | 16,899 | 1,270 |
| Papua New Guinea | 462,840 | 6,732,000 | 16,096 | 2,183 |
| Laos | 236,800 | 6,320,000 | 11,707 | 2,353 |
| Singapore | 710 | 5,183,700 | 307,085 | 52,961 |

# 3. Democracy: Blogger Behavior and Influence

Blogs are seen as extensions of an individual’s online identity. Boyd (2006) characterized blogs as a byproduct of expression and the blogging medium. Blogs are typically classified as a genre, with several sub-genres within the blogosphere. Various genres can be categorized into personal blogs, socio-political blogs, tech blogs, pro blogs, and many more within this medium. Each blog genre determines the type of influence and behavior each blogger exhibits, and the content posted. Genres of blogs also determine the kind of online behavior displayed by the bloggers and their audience, and it also determines the categorization of the blog.

Understanding bloggers' online behavior is necessary since their online approach helps in categorization. Some bloggers focus exclusively on resharing previously created content by amplifying the topic of interest, while others focus on creating a new story. Commenter bloggers share or repost issues they find interesting on their channels. In contrast, the bloggers who make more original content that become the center of discussion are creator bloggers. Many creator bloggers have other social media accounts, such as YouTube and Facebook, to amplify their views (Hopkins, 2014).

## 3.1. Indonesia

Hopkins (2014) Reveals the characteristics of these various bloggers when studying the Indonesian blogosphere; some are based in Indonesia, while others are international. Triastuti (2019) observed the nature of the Indonesian blogosphere and its community modality of operations. Triastuti (2019) found that Indonesian blogger communities operate on one hand as vitias (communities) and, on the other hand, operate as a polis (society) by establishing digital artifacts. These apply to memberships based on locality. This means that each blogging community in Indonesia has a centralized community leader who holds more power than other members. Some notable blogs use administrative or pre-existing geographical regions, e.g., the Bali Blogger community, the Bogor Blogger community, and the Aceh blogger community. Bloggers leading these communities also name their blogs using geographical references. Therefore, creator bloggers in Indonesia leave a trail of geographical artifacts.

Democracy in Indonesia has seen a regular practice due to elections held every five years. Turpyn & Nasucha (2020) studied the role of the public sphere in post-truth democracy in Indonesia. They studied the concept of post-truth in Indonesia through qualitative and quantitative analysis of the public sphere. They classified post-truth as having a lesser impact on what people believe in the digital era as people believe more based on individual emotions or beliefs. They studied the three parties that participated in the 2019 election of Indonesia; they analyzed the comments and reactions to both the incumbent government and the opposition party candidates and arrived at the post-truth perspective of the public summarization below:

1. The incumbent won because of cheating, and the opposition asked to fight the cheating indication. At the same time, the neutral view opined that the elections and presidential elections had been made miserable and fraudulent.
2. The incumbent and the opposition focused their campaign on the current country's conditions.
3. The incumbent is a personal and sensitive satire about the election, while the opposition focuses on hoping for the candidate and the country going forward.

Turpyn & Nasucha (2020) concluded that since in "post-truth," the situation on the ground has no bearing in the influence on the public opinion-forming, individual emotions and beliefs are considered more important. The public sphere increasingly shapes the discourse on fraud allegations in Indonesian democracy. Beginning a debate from opinion will trigger a debate and lead to the opinion appearing as something affecting the public. Although the public sphere benefits from creating an open space for democratic deliberation, the reality on the ground sometimes differs from the truth of the information in the post-truth era on social media. The information in the post-truth era is dominated by certain parties, particularly the ruling party. Compared to the period that predates the post-truth, the public discussion space looks very diverse, and there is no fear of excessive domination of a party.

## 3.2. Malaysia

An individual often creates a blog to voice opinions or publish a medium to drive conversation around the topic of interest. However, with the rise of influential bloggers in public debates, political structures have now infiltrated the blogosphere by creating an online blogging troop called "cybertroop" (Hopkins, 2014). (Mohd Sani & Zengeni, 2010) looked at democratization in Malaysia and the role of Web 2.0 (social media and blogs) in the 2008 general election since social media supports the democratization of knowledge and information. It transforms people from content consumers into content consumers and content producers. With the rise of the web in Malaysia, many corporate media critics, therefore, celebrate the Internet/social media as a naturally fertile ground for independent media, a new media sphere that can compete with corporate media and undermine its influence and authority. Such optimism must be tempered by the realization of corporations already exploiting the Internet to their ends and the challenges independent sites face in gathering resources, establishing credibility, and finding audiences.

In the 2008 general election, despite the Malaysian Communications and Multimedia Commission (MCMC), the Internet in Malaysia flourished despite how regulation stifled the passion of internet users. They went online with the opposition against the government and mainstream media and restrictive campaign rules controlled through ownerships by the ruling Barisan Nasional (BN) government. They turned effectively to blogs, online news portals, and YouTube to dodge a virtual blackout on mainstream media (Mohd Sani & Zengeni, 2010) and (Tarrant, 2008). The world wide web has allowed parties like the DAP, PKR, and PAS to reach voters in their offices and homes, especially young voters between 21 to 40. The Internet is now a player and channel in Malaysian politics, and those who refuse to believe that may have to re-think their views. According to international statistics from the New Straits Times (NST), during the 2008 general election, Malaysia was at a 60% Internet penetration rate. With around 24.8 million citizens, Malaysia had 3.7 million Internet users in 2000. In 2007, this figure was nearly 14 million. This Means Malaysia had a user growth of 302.8% in eight years.

Tapsell (2013), in the study of media movements and the authoritarian electoral regime of Malaysia, observed that many themes and titles are used to describe the Malaysian government regime and how democracy in Southeast Asia view, react and operate toward new media liberalization efforts of various body movements and actors. The 2018 general election was the 14th general election in Malaysia, which was crucial in the country's history due to the blogs' decisive role in the voting pattern. The blogosphere helped bring about the collapse of the ruling National Front (Barisan Nasional) coalition after over 60 years of reign (Nadzri, 2018) and (Abdullah et al., 2019). The 2018 election in Malaysia deviated from analyst predictions of the election result, where the opposition could deny the ruling party the traditional two-thirds parliamentary majority.

## 4. Religion

Religion is one of the levers in cyber activity and online communication in the blogosphere. Bloggers have since used religion as one of the ways to champion democracy and seek election participation in the blogosphere. This is so, as religion is the path of human identity. Since separatist groups have also used religious communication lines to recruit troops, we attempt to see how these religious blogs have been helpful in various campaigns and democracy. The research conducted by (Soriano, n.d.) discusses separatists in the Philippine blogosphere, reviewing how people form information and reinforce beliefs. Many online communities are built around the same ideology; religious bloggers and blogs are fertile ground to consider in this review work. In the context of blogs and bloggers generally attach links to like-minded bloggers and narrow the focus of the debate to their side of political views. (Lovink, 2007) also shares that those bloggers do not encourage debates but validate ideology.

Furthermore, since public opinion on blogs is largely unregulated, there is a thin line between how public opinion could translate into hate and derogatory speech. (Ibrahim, 2009) also argued that since the blogosphere is largely unregulated, expression of passion may also be a form of challenging democratic values: values: her work also looked at the conflictual dimension, its democratic role, and how hate speech functions in subduing the minor group opinion in open contestation. Soriano (n.d.) randomly selected relevant blog threads in the Filipino blogosphere using relevant WordPress, blog posts, and blogs whose theme concentrates on the Muslim separatism issues using Mindanao (which is an island made up of the majority of Filipino Muslims) and the MOA ancestral domain claim as this was subject to hot debate both online and offline media in 2008, Below were the criteria of their analysis sampling selection:

1. Selected threads must come from a blog, excluding organization websites and new websites.
2. The selected thread must also have discussed the Muslim separatist issue within the time of the study.
3. Also, blogs must have had a minimum of three posts from three authors on the subject and must have user interaction.

Soriano (n.d.) argued that the internet provides opportunities to create a public space online. However, offline characteristics from a religious standpoint creep into the public sphere and conclude that the prejudice between Christians and Muslims manifests in public conversation in the blogosphere since stereotyping of the minority Muslim group in the Philippines exists.

## 5. Politics

A book chapter in 2009 by (Ibrahim, 2009) explored how blog synchronizes with offline reality via intertextuality and hypertext. The author describes Intertextuality as the interdependence any literary text has with words preceding it. Intertextuality proclaims a relationship, whether discreet or overt, with other texts. Ibrahim (2009) used the concept of intertextuality in studying the blogosphere in Singapore to understand the dialogism between offline and online discussions. Ibrahim (2009) integrated the concept of intertextuality to explain the role of bloggers in Singapore's democracy. Since a key feature of blogs is hypertext and blogs are made up of linked text, hypertext enables the consideration of networks that blogs can create. Intertextuality online extends offline by relationship on the topic of discussion, the subject of discussion on various blogs is often drawn from the headline of mainstream media. Offline and online narratives play a crucial role in creating intertextuality between offline society and alternative online discussion. Blogs have been able to play a mediating role in Singapore's political landscape since authoritarian rules regulate marked acceptable communication or not acceptable.

### 5.1. The Orange Movement

A thesis by (Nordenson, 2010) described offline action roles in the Kuwait Orange Movement in 2009. Bloggers and activists used the offline interplay to extend online engagement to public discussion outside of online. Undoubtedly, the online and offline synergy was crucial during the Kuwait orange movement. The participant was able to strategize and disseminate news and headlines online, but the offline ensured that voices spread regardless of the internet. Traditionally statistics (Table 2) confirmed by the OpenNet Initiative pointed out that private ISP freely blocks immoral sites. However, the Ministry of communication focuses on blog sites critical of the government or supporting terrorism.

*Table 2: Kuwait, results at a glance, from OpenNet Initiative modified after Nordenson (2010)*

| Filtering evidence | No Suspected Filtering | Selected Filtering | Substantial Filtering | Pervasive Filtering |
|---|---|---|---|---|
| Political | | Yes | | |
| Social Conflict/Security | | Yes | | Yes |
| Internet tool | | | | Yes |

Many Indo-Pacific nations have solid firewalls and surveillance systems for the Internet. Bloggers in Kuwait staged 24 hours no-blogging when surveillance issues reached headlines in the fall of 2009, but orange movement bloggers could still express their views almost unhindered (Nordenson, 2010). With the adoption of blogs, Internet penetration, and the Orange Movement, some conversation around using blogs as a political activism tool to get women to vote in Kuwait was mentioned among the youth. Aday et al. (2010) pointed out that bloggers have played a key role in mobilizing contentious politics.

However, he argues that this is a relatively new phenomenon in Kuwait, originating with the Orange Movement: “The Kuwaiti case is fascinating since, before 2006, most observers had seen the Kuwaiti blogosphere as relatively disengaged from politics and marginal to the public realm” (Nordenson, 2010).

The Orange Movement became a focal point to prevent an additional five districts in Kuwait. The reformer group’s underlying reason for the demand for additional districts was vote-buying and clientelism (also known as Service members in Kuwait). This is because it will require a small number of a vote to win an election.

To trace the role of the Orange Movement (Nordenson, 2010), take a closer look at how the Kuwait blogosphere started around 2003 with few communities that were not political. With Kuwait, constitutional issues grew during the spring of 2006 is one of the reasons why the orange movement’s involvement in events like “Ayya” of “The Ultimate” speaks of being “fed up.” These activities started from the blog, although the idea was not to launch a movement but to blog about what was happening. Kuwait bloggers were able to organize politics because many of them existed around the same time through blog aggregators, helping bloggers team up for the orange movement as the blogosphere in Kuwait has laid the foundation. The orange movement activity in Kuwait was able to use the blogs and bloggers as the reference point in reaching out to other offline pressure groups whom politicians and other prominent Kuwaitis have previously arranged to support their cause for an event or gathering called “The re-conquest of a nation” which was mobilized by a blogger the “Kuwaitjunior” blog. The Orange Movement pressured the government to reduce the electoral district to five instead of twenty-five, a tactic by people in power to win elections easily.

Deconstructing the politics that influenced the “Malay Vote” in 2018, (Rahman, 2018) unraveled the electoral revolt against the incumbent Prime Minister Najib Razak. The findings by (Rahman, 2018) showed that Malaysian voters could vote for the then-opposition due to various economic and socio-political factors, thereby driving the voters against the ruling Barisan Nasional coalition. These voters’ demography was divided along the path of the urban-rural divide, and what issues were dominant that affected the voting decision? Rural areas play a crucial role in Malaysian political study. They occupy large geographical areas with small populations.

Still, constituents are not ethnically mixed compared to the urban areas, where constituents are more ethnically mixed due to migration and mixed political ideology. So rural issues were usually considered Malay issues because of the nature of constituents in the rural region. Although the government at the center can consolidate power using racialization, they can depend on loyalty at the rural level. The Malaysian rural community’s view of politics is different from the American rural community, which is judged to be driven by cultural beliefs and emotions. The blogosphere may not have a way of dividing information between the rural and the urban communities in Malaysia, but the means of information dissemination and access to information differs. Therefore, information deemed true in the urban region may be fake news in the rural community. The election tipped in favor of the opposition party for different reasons at the urban and rural levels. At the urban level, it was through the awareness of the principle of equality and social justice. At the rural level, it was more about inequality that comes from patronage politics.

Further study on Malaysia politics by Leong (n.d.) demonstrated how in Malaysia, the Barisan Nasional (BN) uses controlled media via licensing and censorship, like many other Asian countries, to maintain political power and wipe e-mention by pushing various narratives through traditional media. However, (Leong, 2017) also studied how the new media play a role. In challenging the status quo in Malaysia.

An example was how the new media played a key role during the 2008 Malaysia election tsunami that eroded the popularity of the Barisan Nasional. The blog expanded the public sphere in Malaysia and the democratic process to involve many citizens by im- proving information dissemination, sourcing for funding, and mobilization. Blogs and social media have been categorized as the fifth estate since the emergence of Web 2.0 as they can pressure government and political office holders. However, the new media, specifically Web 2.0 (blogs), cannot change the government narratives without moving the conversation offline despite allowing many voices to emerge. The blogs and new media were successfully used in Malaysia, but the government has since joined these spaces to dominate and normalize the internet. The Internet has helped galvanize the otherwise silent voice since the traditional gatekeepers can now be bypassed.

## 5.2. The Philippine Blogosphere

The Philippine blogosphere (DUAQUI, 2012) explored web blogs' political voices. Table 3 shows the internet growth usage in the Philippines between 2000-2010. The Philippian blogosphere acts as a vibrant public space and opinion marker despite internet penetration due to the strong readership culture within the Filipino and sustained efforts in writing blogs. DUAQUI (2012) explored various themes discussed in the Philippines blogosphere and arrived at five different themes ranked in other of their frequencies, with the theme "Role of the Philippine executive" being the highest, mainly referring to blogs tackling the critical accomplishment of the Philippine presidents Figure 4 and Figure 5. The subject and pattern of each theme focus on various aspects of the country's democracy that these blog sites are tackling. For example, the blogs under the theme of "Internet Politics" tackle issues on the role of the internet in Philippine democracy, e.g., blogs, Facebook, and other social media platforms.

Furthermore, (DUAQUI, 2012) looked at the most commented themes within the Philippine blogosphere and presented the top five heavily commented themes in Table 4. DUAQUI (2012) also explored the issues directly related to democracy; the role of the Philippine executive significantly gathered more frequency than other debates like political change, women's suffrage, and corruption, which is the next closest issue in the Philippine blogosphere that received attention like the role of the Philippines executives. DUAQUI (2012) claimed the internet to be the "Internet as a democratizer," reaching this conclusion due to various data collected in the Philippines blogosphere; since the Philippines home and abroad can participate in democracy through virtual resources published online on various blogs.

Pooi Yin Leong (2019) discussed political communication in Malaysia and the use of new media in the Malaysian public sphere, public sphere political deliberation in a liberal democratic structure provides citizens with a platform for choosing their political representatives. Compared to the media in Malaysia, the media is used for control and bias against the opposing parties. Pooi Yin Leong (2019) highlighted opposing voices as early adopters of new media in Web 2.0. The interactivity of Web 2.0 provides everyone with a platform for deliberation, and this allows opposition parties to circumvent tight control over traditional media and challenge the official narrative. The success of the new media and platform that Web 2.0 and the blogosphere provide challenges to the traditional grip on political news and narrative dissemination.

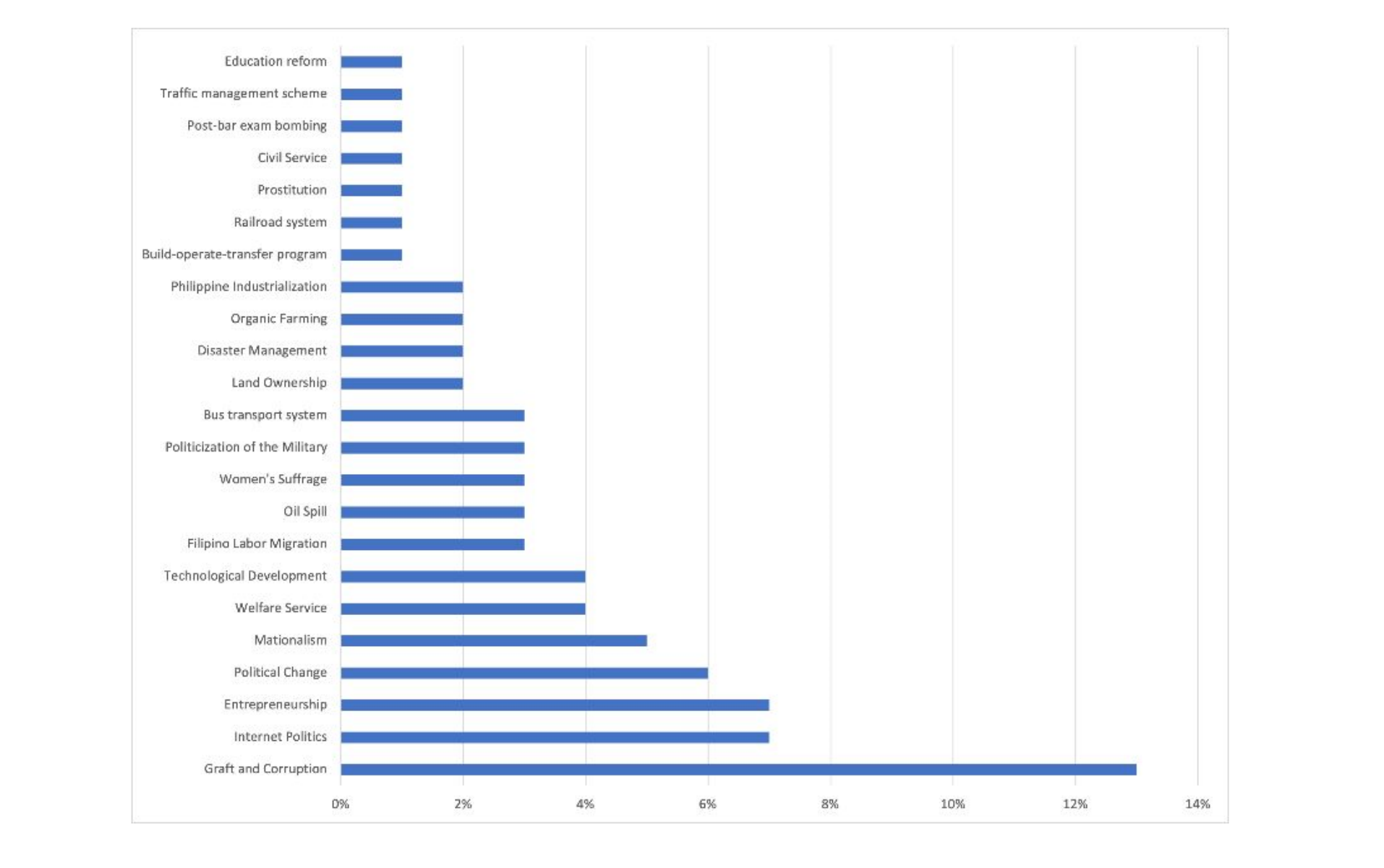

*Figure 4: Showing Composition of blog themes studied by DUAQUI (2012).*

*Table 3: Growth of Internet users in the Philippines, 2000-2010 modified after DUAQUI (2012). Source: http://www.internetworldstats.com/asia/ph.htm*

| Year | Internet Users | Population | % of Population | Source |
|---|---|---|---|---|
| 2000 | 2,000,000 | 78,181,900 | 2.6 | ITU |
| 2005 | 7,820,000 | 84,174,092 | 9.3 | C.I. Almanac |
| 2008 | 14,000,000 | 96,061,683 | 14.6 | Yahoo |
| 2009 | 24,000,000 | 97,976,603 | 24.5 | Nielsen |
| 2010 | 29,700,000 | 99,900,177 | 29.7 | ITU |

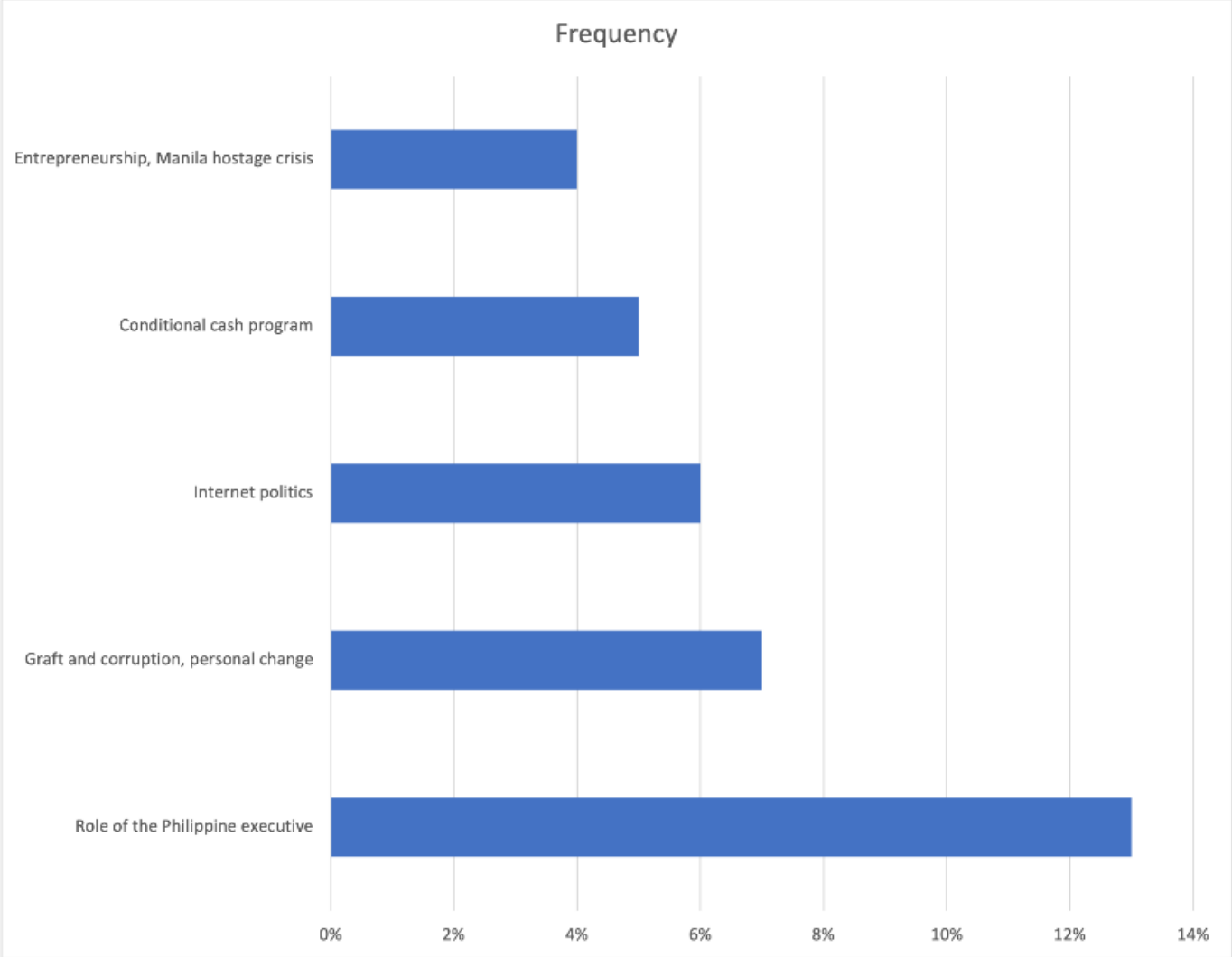

*Figure 5: Top five themes discussed in the Philippine blogosphere, DUAQUI (2012).*

*Table 4: Top five heavily commented themes in the Philippine Blogosphere DUAQUI (2012).*

| S/No | Theme(s) | Number of Comments |
|---|---|---|
| 1 | Personal change | 221 |
| 2 | Population management | 177 |
| 3 | Education reform | 164 |
| 4 | Manila hostage crisis | 149 |
| 5 | Internet politics | 84 |

### 5.3. Australian Blogosphere and Online Deliberation

Mummery and Rodan (2013) compared three very active blogs in the Australian blogosphere whose content focuses on politics and deliberative democracy. These blogs are Larvatus Prodeo, Bartlett's Blog, and Bolt's Blog. The Author compared the blogging rate and threads with most activities to study how deliberative politics in the public sphere works. Each blog showed a different engagement pattern, with Blot's blog having a proportionate pattern showing that two-thirds of the participants posted just once compared to Larvatus Prodeo and Bartlett's Blog, which enjoyed an average of three comments on a post Table 5.

Although Bartlett's blog increases the number of posts, the interaction pattern on these blogs may have been affected by the moderation policies since these blogs have comment policies in place. These deliberation policies align with Habermas's view that online deliberation in the public sphere should require moderation measures for check and balance. Mummery and Rodan (2013) found that within these three tops, Australian political blogs' Truth claims are left open and uncontested.

Furthermore, (Mummery & Rodan, 2013) monitored the role of various blogs covering elections and discussing politics during the 2010 election in Australia by considering three politically focused blogs in Australia: Larvatus Prodeo, Andrew Bolt, and Andrew Bartlett. These three studied blogs are of various styles and compositions, considering how they blog. The Larvatus Prodeo is a group-authored blog, Andrew Bolt's blog is a sole-authored blog (conservative commentator), and Andrew Bartlett's blog sole-authored blog by a former Senator. From the thread studied by (Mummery and Rodan, 2013) in Larvatus Prodeo blogs, a thread: Why I'm voting for the Greens tomorrow by Mark Bahnisch received 180 comments in 19 hours 26 minutes between 13 August 20 and August 21, 2010. The study's key findings are that the authors in the Larvatus Prodeo blogs received regular comments since these authors share interests and are knowledgeable about political issues. Secondly, these authors have a common goal of talking about issues.

## 6. Identifying Types of Bloggers

The key to identifying the various type of bloggers involves considering their mode of operation and how they interact, their motives, and their intentions in the blogosphere. To help readers understand these blogs, this work identifies three main groups of bloggers, while this is partial. It does cover the key characteristics of the important type of bloggers in the Indo-Pacific region and how they influence democracy in this region. In the next section, we dive into these various types of bloggers and their role in democracy in the region.

### 6.1. Cybertrooper Bloggers

Historically, countries and governments recruit troops to defend the country against internal and external aggression. With the advent of cyberspace, the blogosphere is a micro-nucleus of cyberspace. The blogosphere has given birth to public media controlled and owned by either personal or corporate opinion (media house owning a blog). Political organizations, countries, and governments have entered the blogosphere by recruiting troops online for blogging activities. Many of these troops are either recruited to champion narratives or opinions. Cybertroop plays an active role online, helping the organization that recruits them in countering an opinion against their paid organization. Cybertroopers are not limited to bloggers. They are troops engaged or involved in cyberspace to counter or push a narrative. Cybertroopers help manipulates online opinion and presence by engaging on behalf of their benefactors, but authoritarian regimes have adopted this approach to participate in democracies worldwide. Malaysia is among the countries where manipulation through social media remains a growing threat to democracy. Cybertroopers originally found their roots in Malaysia but have since seen worldwide adoption for the term as someone engaged in disseminating political propaganda on the Internet.

During the Malaysian election study, the concept of cybertroopers' roles in elections was first discussed by (Hopkins, 2014). Hopkins (2014) attempts to explore how the cybertroopers engaged by the government suppress opposing views online and spread the government's view. Mohamed et al. (2011) Define cybertroopers as people who carry out harassment in the virtual space. One of the characteristics of cybertroopers, as listed by (Mohamed et al., 2011), is that they can argue well, are well-educated, and possess good language skills in Bahasa Melayu and English. Cybertroopers can also defend the ruling party while attacking the opposition parties rationally. The ruling party can warn off attacks using cybertrooper agents.

Johns and Cheong (2019) considered cybertroopers and how state authorities censor online mass movements of citizens using the theories of "networked affected" and "affective publics" affecting Malaysian digital citizens. With the Malaysian mainstream media controlled by the Barisan Nasional government, the opposition in Malaysia relied on various digital platforms like blogs to mobilize support. The Barisan Nasional government responded with consistent anti-Bersih messages created by the employed cybertroopers sympathetic to the Barisan Nasional government at the center. The Barisan Nasional government in Malaysia also employed Cambridge Analytica in its 2013 election campaign to win the key battleground seat of Kedah (Johns and Cheong, 2019).

The role of cybertroopers during the Malaysian general election in 2014 was discussed by (Hopkins, 2014). Hopkins (2014) observed that a few months after the election. Despite barbs from the United Malays National Organization cybertroopers, deep divisions among Malaysian voters seem to have escalated. People who voted questioned Mahathir's intentions and actions, while generational United Malays National Organization supporters lamented their spur-of-the-moment vote for the opposition. United Malays National Organization stalwarts at the grassroots level imagined new ways to position themselves to benefit from expected patronage under new guises. Many felt that the Pakatan Harapan takeover would be a temporary diversion and that United Malays National Organization would soon return to power. This ground sentiment was reflected in a statement by Ahmad Zahid Hamidi in the lead-up to the United Malays National Organization presidential election. Despite some urban Malaysians' reluctance to back ethnocentric politics or policies, their rural counterparts appear eager to see the new government fall short of its promises.

The work of (Barendregt and Schneider, 2020) categorized the cybertroopers as fake news peddlers the government employed to counter the Bersih protesters who took to Malaysian large city streets during the 2015 protest. The cybertrooper behavior was also described as trolls and producers of fake news. Since digital activism provides a pathway for cybertroopers to cooperate within the same public sphere allows both to co-exist when variety in opinion matters. Cybertroopers in the Indo-Pacific can hide under digital activism to push government narratives. Malaysia's government adopted a communication strategy in the wake of the voter tsunami in the 2008 and 2013 elections. And This strategy believed that since the pro-opposition group dominated the online media, the Barisan National government believed in having a countermeasure to tackle online opposition (Cheong, 2020). The counterstrategy means that the Barisan National had adopted the cybertrooper as one of its information arms, particularly with blogs playing crucial roles in opposition to political gains.

#### 6.1.1. Difference Between Cybertrooper and Citizen Journalism

Although, some authors have compared citizen journalism in the blogosphere to the activities of keyboard warriors and cybertroopers. Mahamed et al. (2020) examined the differences between citizen journalism and cybertroopers and how citizen journalism and participatory journalism promote democracy within a nation. First is the recklessness of cybertroopers who share what they believe to be true because they have no sense of responsibility. Secondly, citizen journalism creates a positive environment for conversation compared to cybertroopers and keyboard warriors. Citizen journalism does not understand ethics when sharing information online compared to employed cybertroopers and keyboard warriors.

Also, (Mahamed et al., 2020) highlight key challenges citizen journalists face. These are risks of misconception from the public that is compassionate toward keyboard warriors and cybertroopers risk of cyberbullying by cybertroopers and keyboard warriors. Mahamed et al. (2020) finally believed that citizen journalists, mainly the youth, use the online medium to express 19 themselves but are usually counter-twisted by the cybertroopers. According to Authors in Qiu (1999), The Chinese authoritarian regime uses the media as one of the strategies to diffuse the public sphere since it is a way of implementing virtual censorship to reduce the effects of public sphere deliberation so that the citizen will communicate in mediums like blogs in the old hegemonic way.

#### 6.1.2. Similarity Between Cybertrooper and Social Bots

Since cybertroopers behave like bots, they can also be seen as human bots in the study of social network analysis. Khaund et al. (2022) surveyed social bots and their online coordination during campaigns. Their work suggested it is important to leverage existing or known instances of coordination from previous literature and develop a network measure-based assessment framework. Since these Bots play an active role in content dissemination, it is important to note that actual human users monitor these accounts. Indicators based on resource sharing, such as identical texts, URLs, etc., at the same time, along with concrete network science theories, will help us identify patterns to study coordination Khaund et al. (2021). Cybertrooper has a similar pattern in this region, especially in countries where authoritarian control has existed historically.

### 6.2. Commentator Bloggers

Traditional audiences have also begun to blog due to the affordability of commenting on existing blog posts. This ease of access is termed "personalization affordance" by Hopkins (2013), as readers have now developed a sense of the blogger. Readers can communicate and extend the post's content in a blog post. Hyperlink sharing also develops interpersonal interaction in comments, and sustained interaction usually forms the basis for blogging. This was also demonstrated during the 2008 Malaysian election period, according to Hopkins (2013). The effectiveness of political conversation using blogs has helped many audiences who are regular commenters to become politically active themselves; this is clearly stated according to Hakim, Corbitt, and Young (n.d.). Commentator bloggers have also used blogs to assist at election locations by providing information Hakim et al. (n.d.).

In Australia, Mummery and Rodan (2013) showed an example of commentator bloggers where a group-authored blog discusses various political topics extensively and can critique and some- times hyperlink to various materials from various news media. These commentator bloggers are very knowledgeable about discussion and have a keen interest in various areas they comment about in various blog threads. The exchange pattern by commentator bloggers usually fits in the deliberation of democracy, provided these conversations elaborate on their viewpoints and respond to others' points.

During the political tsunami from 1998 to 2008 in Malaysia, the role of bloggers and commentator bloggers in regulating cyber political activism was documented by Salleh (2013). The political bloggers and commentators were able to warn take down or counter comments that were either insulting to others' views or not seen as authentic since the blogs were used as a participatory medium Salleh (2013). Commentator bloggers also played a crucial role in elections; many have won political seats. An example was during the political tsunami in Malaysia, with five bloggers getting parliament seats in the 2008 election Khoo (2016). In comparison to Malaysia, the government pays some commentator bloggers to lead discussions. According to Abbott, Macdonald, and Givens (2013), In China, some group commentators are called Fifty Cent Party, and The government pays fifty cents to lead the discussion in a pro-regime direction. The fifty-cent party memberships are estimated to be about 280,000 as of 2008 Abbott et al. (2013). Commentator bloggers have also emerged as thought leaders promoted by others within the community since members can contest ideas quickly Tang (2009).
The role of commentator bloggers in American politics has been categorized into two categories, namely "agenda-setting" and "agenda seekers," according to Davis (2009). This is because they can set the conversation's tone and center the discussion by reacting to information from established media. Commentator bloggers' agenda-setting is more pronounced during the election period in American politics. However, commentator bloggers are still really on traditional news media to set the agenda since the bloggers' efforts are usually reactionary to the news headlines.

## 6.3. Congressional Bloggers

Congressional blogging refers to bloggers who are senate or people's representatives' politics that use blogs and websites to take credit or provide updates on various issues they have either supported on the floor or self-credits for achievements. According to (Pole, 2006), one of the characteristics of congressional blogs is comment sections which facilitate communication between the members and their constituents. The study uses content analysis and demographic study to find who are the congressional members that actively participate in blogging and answer questions like what their educational background is, race, why they blog, and how are their followers participating in democracy through the blogging comment section as a feedback mechanism to the congress members.

In work published by (Pole, 2006), content analysis of selected congressional blogs was studied. Among the congressional blogs studied, the demographic data showed that the mean age of the congressional bloggers is compared to the traditional blogging pattern over the world, where bloggers are more millennials who were at the center of the adoption of web 2.0. Pole (2006) also found that of the congressional bloggers, 80 percent were males and more white congressmen and women compared to the black congress members. Her findings also show that white congress member bloggers are more educated than other races. These congressmen actively use the blog to advertise their work, take positions on issues, and inform their constituents. An example of her work is the Majority Leader John Boehner, who was able to use blogging as a credit-claiming mechanism through blog posts that demonstrate him as the party leader, taking credit for all the Republican party did at the time. This pattern of position-taking and the credit-claiming system was widely observed across the party line.

The adoption of blogs by congress members and political actors in the American blogosphere has also seen prominent politicians hire political bloggers to grow their blog presence. Kaye (2010) work demonstrated an example of prominent politicians and ex-congressmen who adopted blogging in the mid-term 2008 election and during the US presidential election with Hillary Clinton, an example of a prominent politician who had hired political bloggers to build her online image. The Obama campaign team also used blogs or websites to put content online. Kaye (2010) reported that BarackObama.com had posted about 1800 videos on his blogs which gathered cumulative 18 million views. Politicians in America extensively use blogs for advertising their manifesto and work while also taking credit for various bills and implementation and antagonizing views of the opposing party. Kaye (2010) statistics of political party affiliation agree with (Pole, 2006) that the Republican party dominates the blogging community among the congressional participants. The blog began having a strong influence in the American election with heavy usage from 2004 to date. Presidential candidates like George W. Bush and Senator John F. Kerry hosted weblogs on their official websites. During the 2004 campaign, Bush blogs pushed out an average of 8 posts per day during the election period (Trammell, 2006).

*Table: Showing political blogs in Australia with thread statistics after [21]*

| Blog Thread | Thread | Number of bloggers in comments thread | Number of bloggers who post just once | Percentage of bloggers posting once | Percentage of bloggers posting more than once | Average number of posts for blogger posting more than once |
|---|---|---|---|---|---|---|
| Larvatus Prodeo | Why I'm voting for the Greens tomorrow | 180 | 44 | 56.41 | 43.59 | 4.00 |
| Larvatus Prodeo | Taking our time | 172 | 32 | 52.46 | 47.54 | 4.82 |
| Bolt's Blog | Into the second week of campaigning | 51 | 5 | 45.46 | 54.54 | 7.67 |
| Bolt's Blog | Congratulations Prime Minister whoever | 238 | 115 | 72.78 | 27.22 | 2.86 |
| Bolt's Blog | Congratulations Prime Minister whoever | 532 | 191 | 64.09 | 35.91 | 3.19 |

# 7. Results and Analysis

This section discusses various analysis and findings from the collected journals used in this survey. The section is grouped into bibliography representation analysis, publication trend, publication type, keyword analysis, network analysis of authors, and findings.

## 7.1. Bibliography Representation and Analysis

This section presents an analysis of the bibliography used in this study. This survey involved looking at the trends, keywords, common themes, and popular words that appear consistently throughout the various publications used in this study. The work aims to capture relevant information and themes within the various studies and prose the recognition of the dominant theme in defining this work to capture the study area succinctly.

## 7.2. Publication Trend

To observe trends in literature, 225 publications since 2003 were collected using various data collection methods by querying google scholar, web of science, and other indexing platforms. We apply the query of "Indo-Pacific" and some other country names along with "blogs" as querying parameters. We choose to expand the year of data collection to allow capturing information since the work focuses on a region and blogs. Our study found that the interest in publications around countries belonging to the Indo-Pacific region peaks around when there are election activities or events against ruling governments. This could be linked to the fact that more data are available when such high-impact events occur example is, the peak between 2007-2009 is pre and post-elections in countries like Malaysia and in 2010 in the Philippines. Similarly, the US also had elections which corroborate our findings that blogging activities by congressional bloggers in a country like the US were on the rise in 2008, and cybertroopers in other Indo-Pacific regions at the time of elections and other high-impact events.

## 7.3. Publication Type

We also grouped the various publications' bibliographies exported from our reference manager library to know which publications contributed most to our study. We observed that journal articles contributed significantly to our study, with 178 articles followed by documents which are policy documents published by various agencies. The next most con- tributing publications are followed by book sections and books along with conference papers, as shown in Figure 6.

## 7.4. Keywords Analysis of Bibliography

Our work also applies a further in-depth analysis of the various keywords used by the authors of the 225 publications studied. We extract these keywords or keyphrases from the titles, keywords, manual tags, and auto-tags presented by the journal indexing platforms like google scholar. This allows us to see which keywords are dominant and help us align these keywords to the survey body. The most dominant key- words we found are as shown in Figure 7 with words like "Indo," "Social Media," "Pacific Strategy," "Barisan Nasional," and "Virtual Space," to list a few having more visibility in various publications than the others. We also observed that many of these publications have consistency in using words that are synonymous with how citizens contribute online to the blogosphere, particularly words like "online," "virtual space," and "citizen journalism" could be exchanged in the blogosphere. Since it also means an online community where citizens can have their own media.

Furthermore, we also use a word cloud to visualize common words across publication titles to help us determine which word is common in most of the collected articles. We found phrases like "Deliberative Democracy," "Communication," "Blog," "Social Media," and others like accountability. This set of phrases signifies the participatory community and the interest of citizens to use blogs and other social media tools to discuss and hold regimes accountable by making demands and applying pressure on governments Figures 8, 9, and 10.

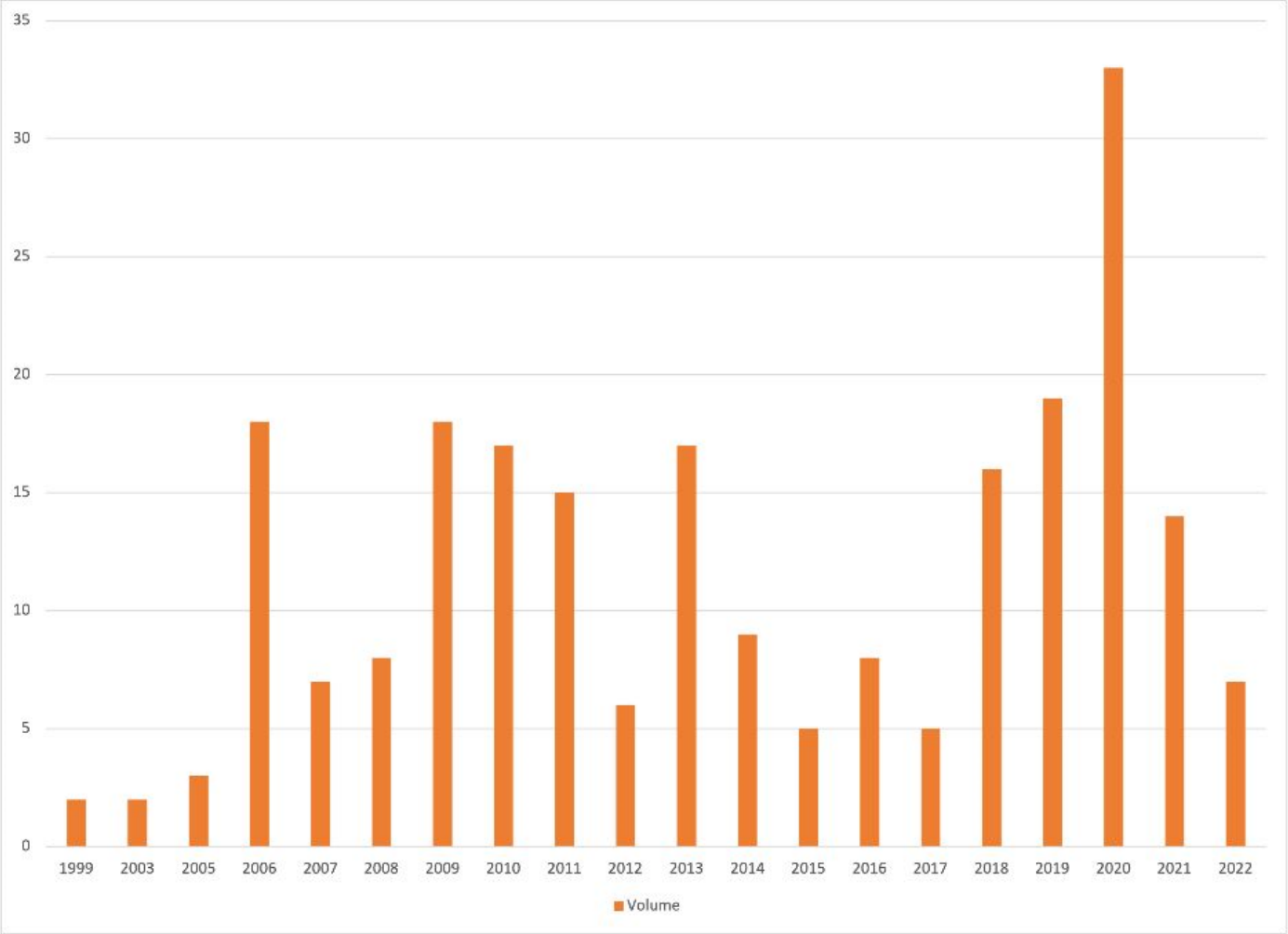

*Figure 6: Showing the volume of publications over the years in our selected bibliography*

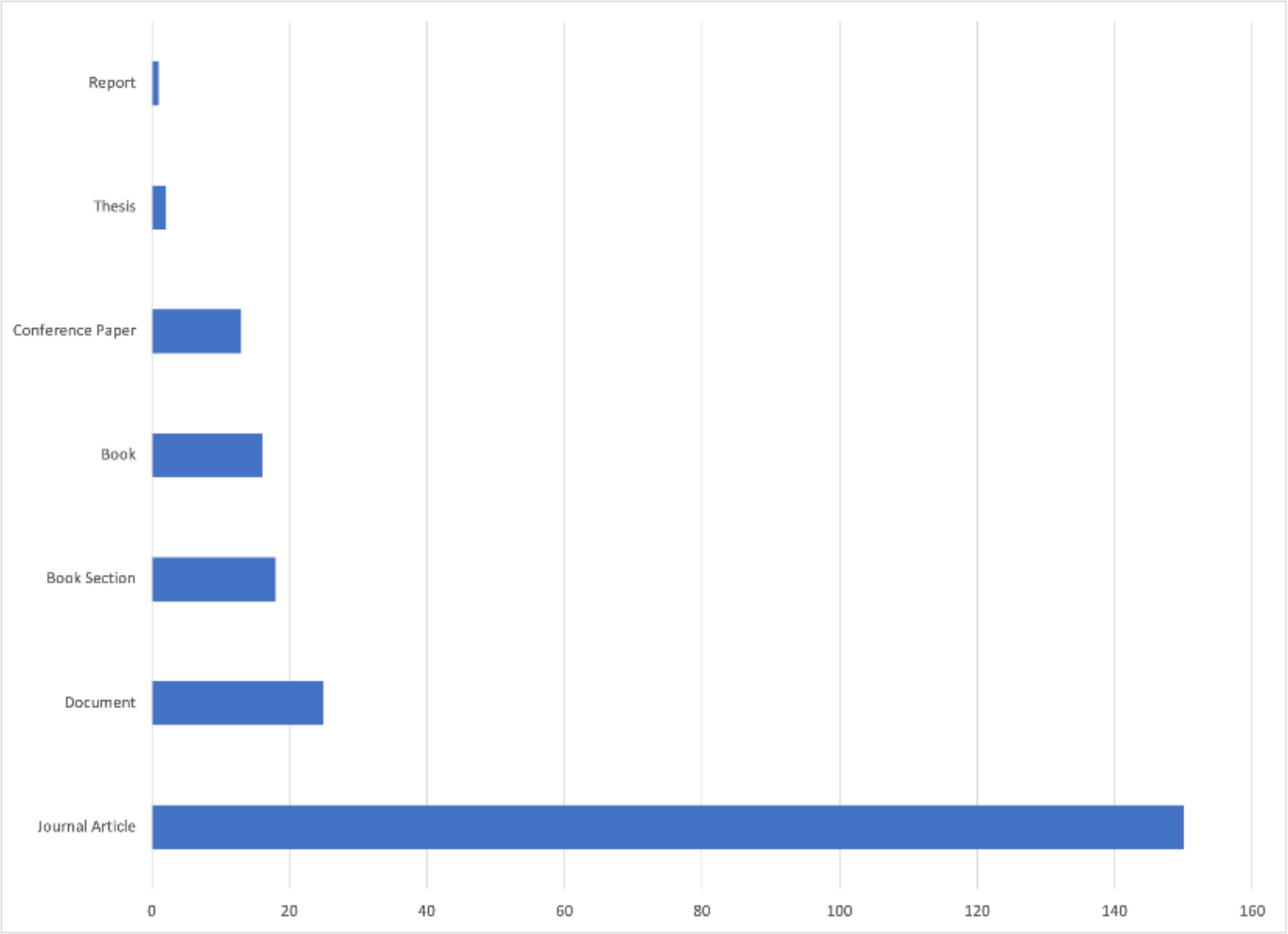

*Figure 7: Showing publication type category to help readers understand the composition of our selected bibliography*

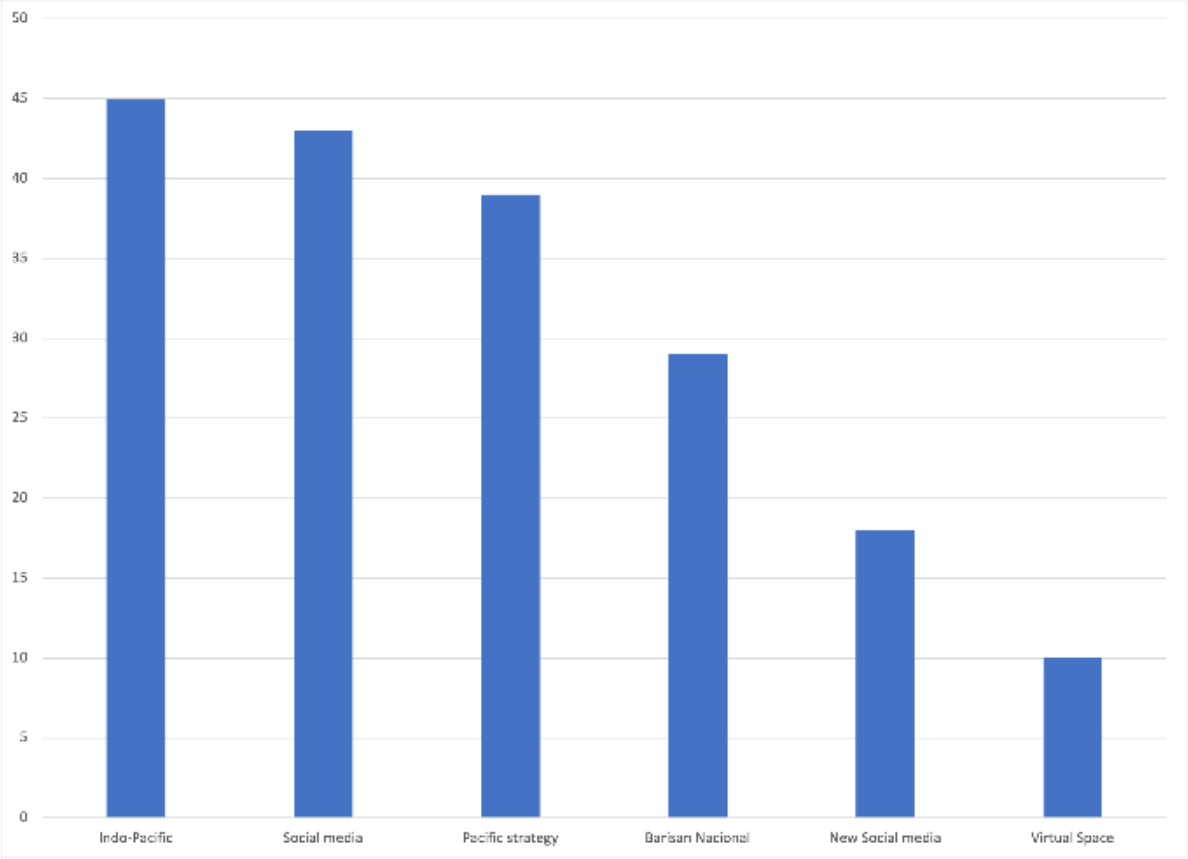

*Figure 8: Showing frequent keywords in various publication titles of the selected bibliography*

## 7.5. Network of Author Analysis

The network analysis of authors and citation themes are shown in Figures 11, 12, 13, 14, and 15. We used the vosviewer to visualize and analyze the network of co-authors. Our selected bibliography has 13 clusters of networks of the larger network group. The clusters of co-authors also have clusters with a few documents greater than five between them. Figure 11 shows cluster 1 with the most author presented with authors. These clusters were obtained using the Association strength normalization method; our visualization uses the number of documents as the weight. The network in Figure 11 is an undirected network showing how the authors with the highest number of papers participated along with a path to other authors in the network. The network in Figures 11, 12, 13, 14, and 15 can be combined to show researchers and the themes in which they participated. Their contribution to the various body of work in the indo- pacific region is also shown in Figures 13 and 15, how these authors started studying blogs and subsequent research areas shifted focus to deliberative democracy and internet participation of citizens in their countries' democracy.

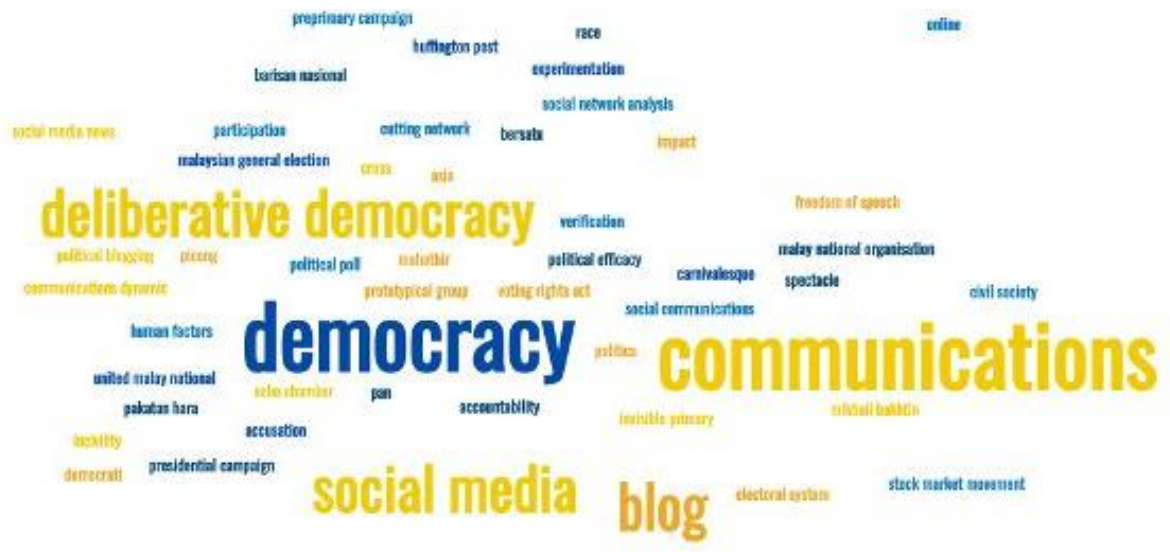

*Figure 9: Showing word cloud of key themes in various publications titles*

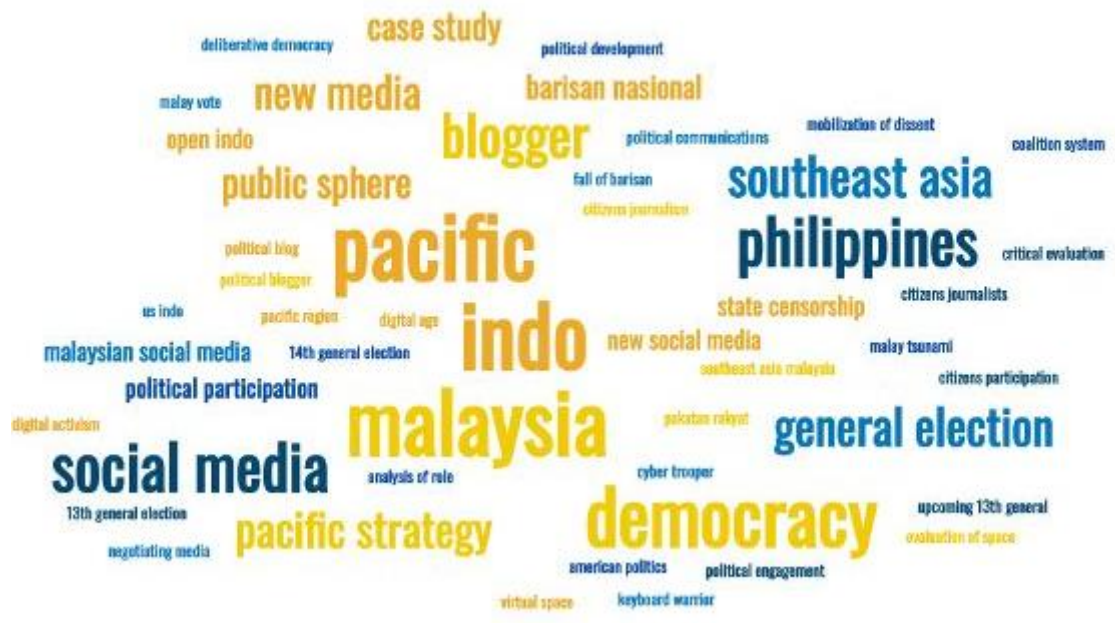

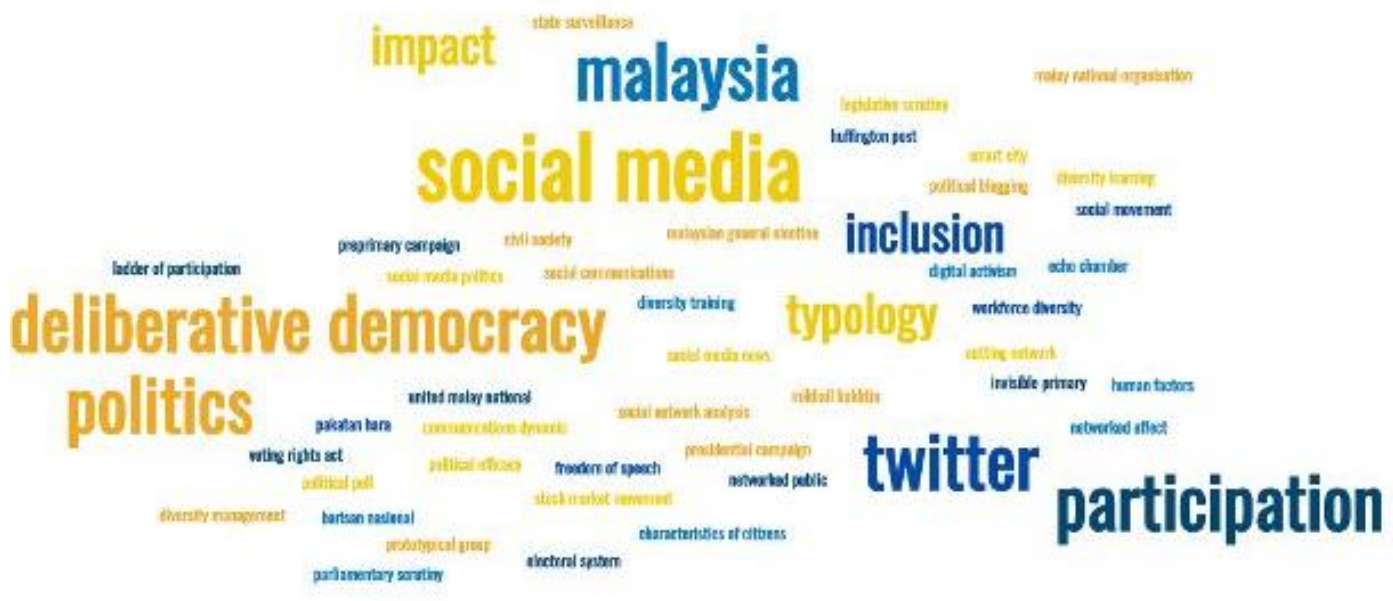

*Figure 11: Showing word cloud for auto tags extracted from selected bibliography.*

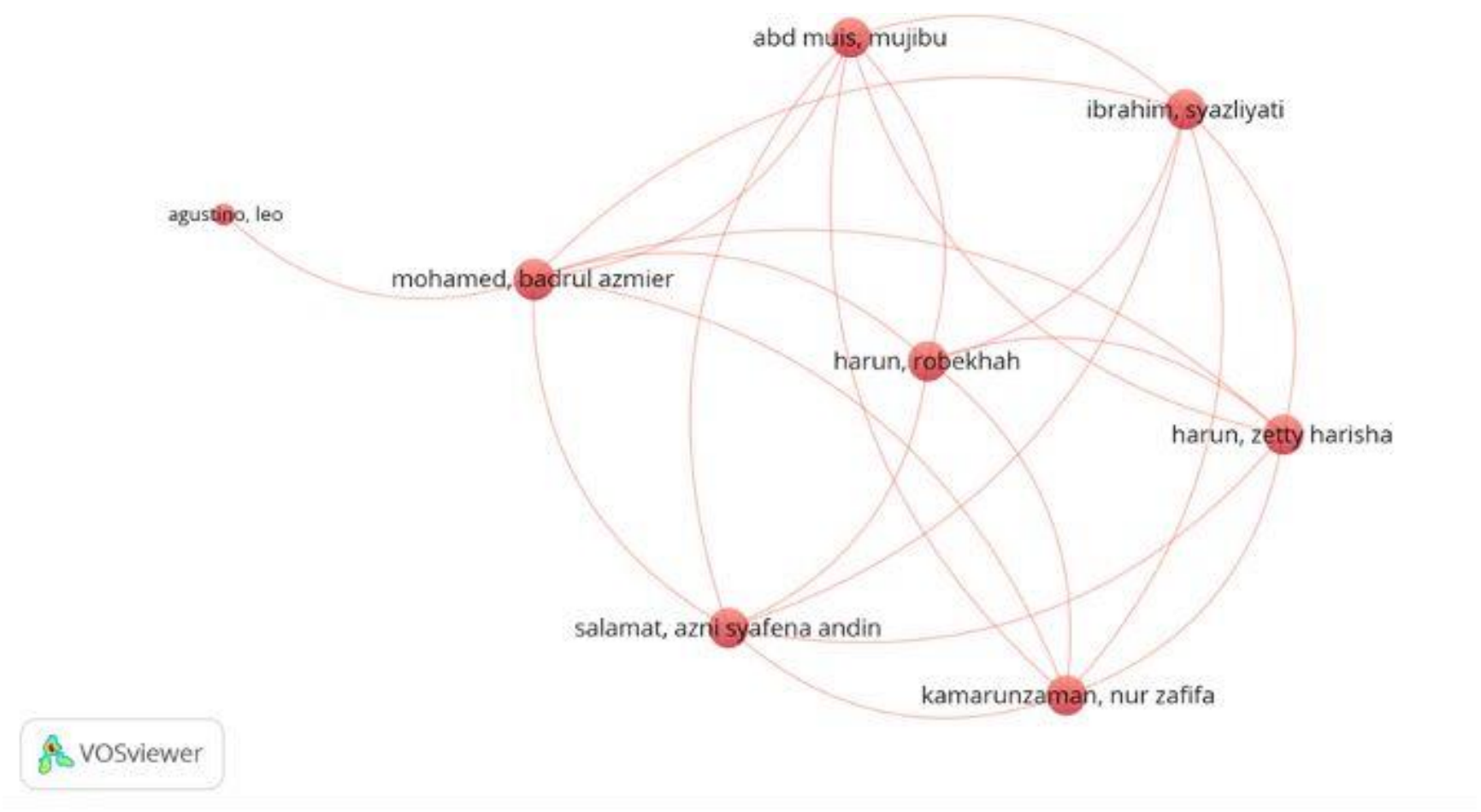

*Figure 12: Showing Cluster 1 of the Author-Network.*

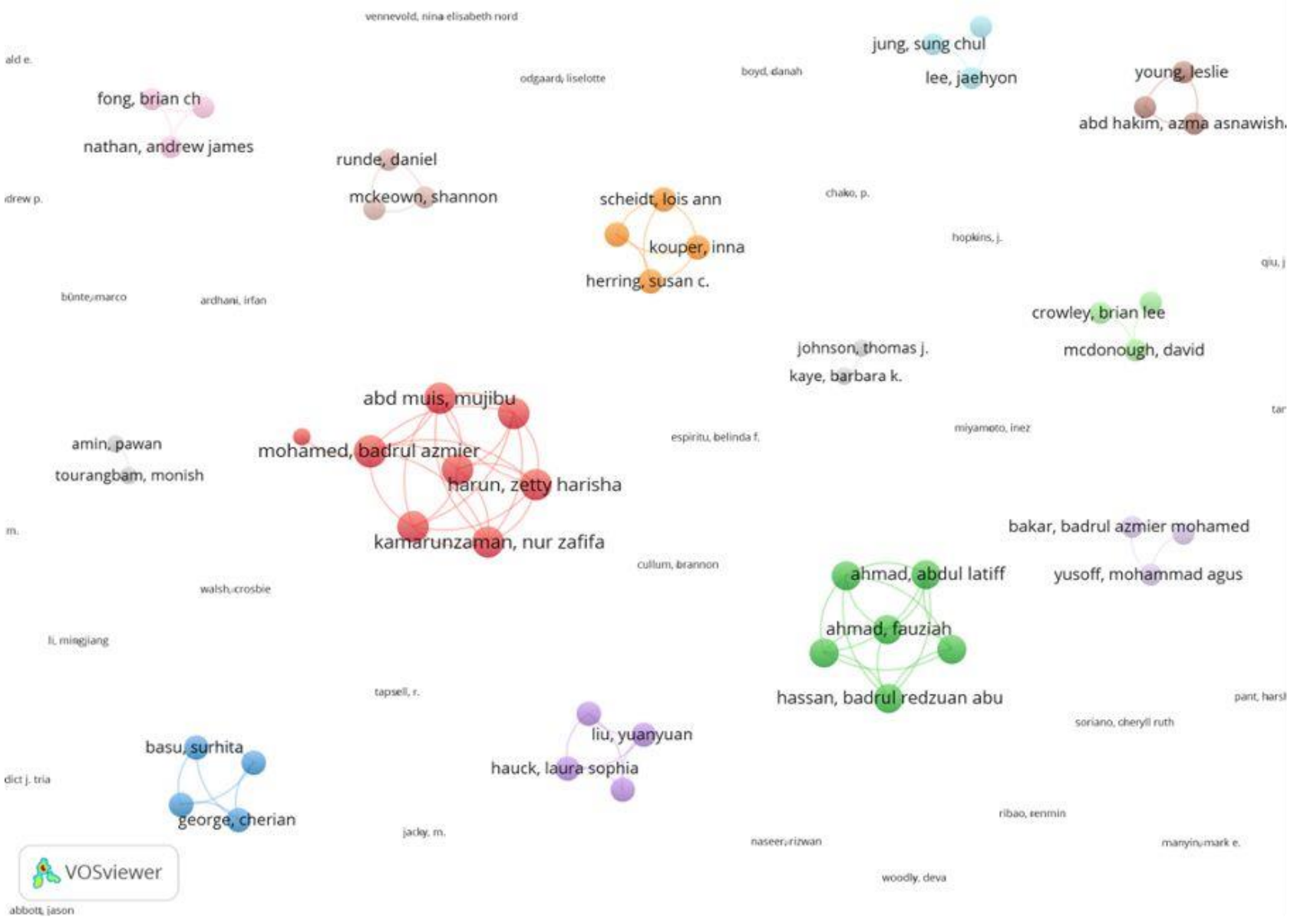

*Figure 13: Showing author network from our selected bibliography and the top clusters.*

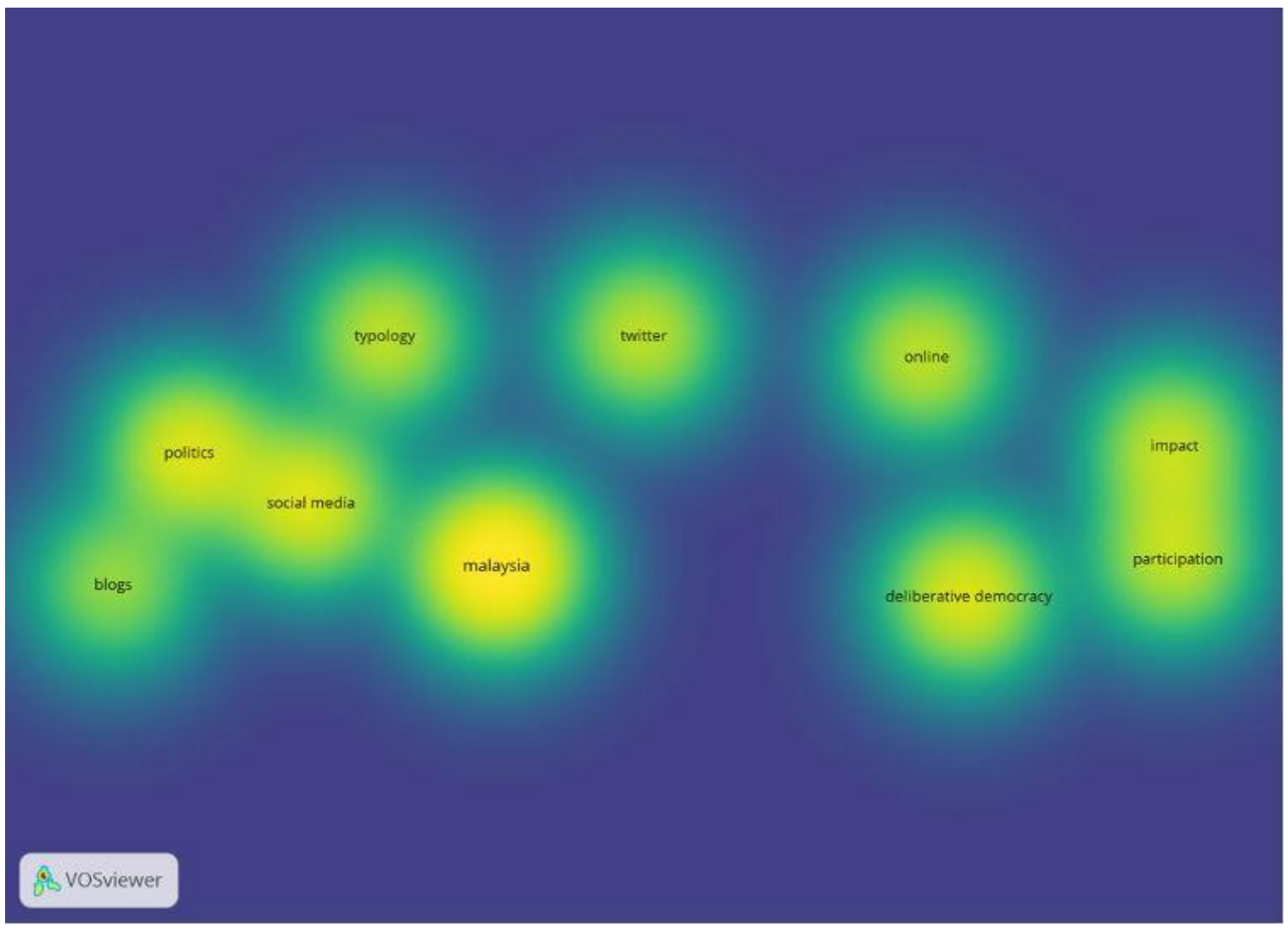

*Figure 14: Key theme density network visualization showing key phrases by the selected bibliographic.*

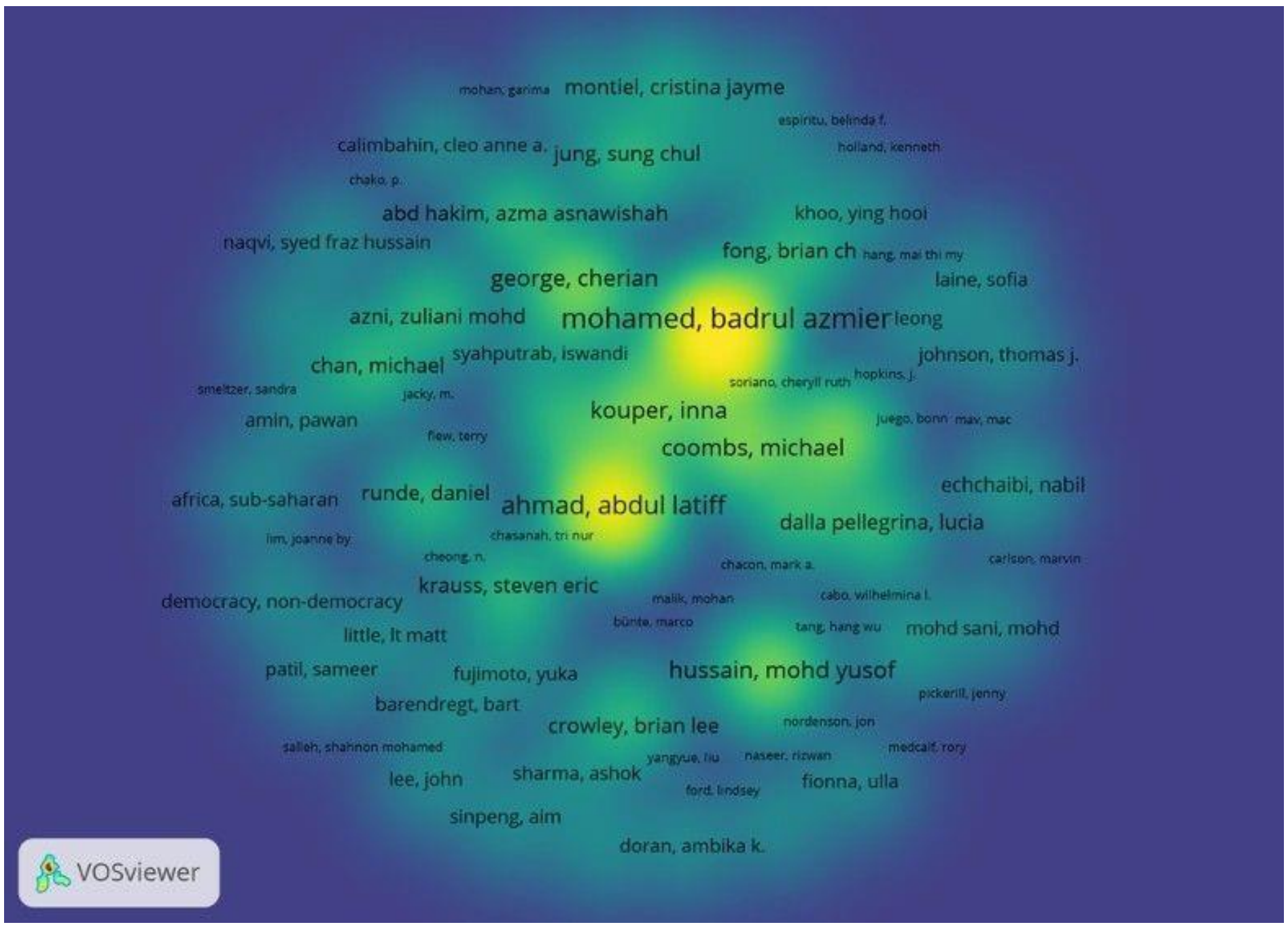

*Figure 15: Co-Author density visualization of selected bibliographic.*

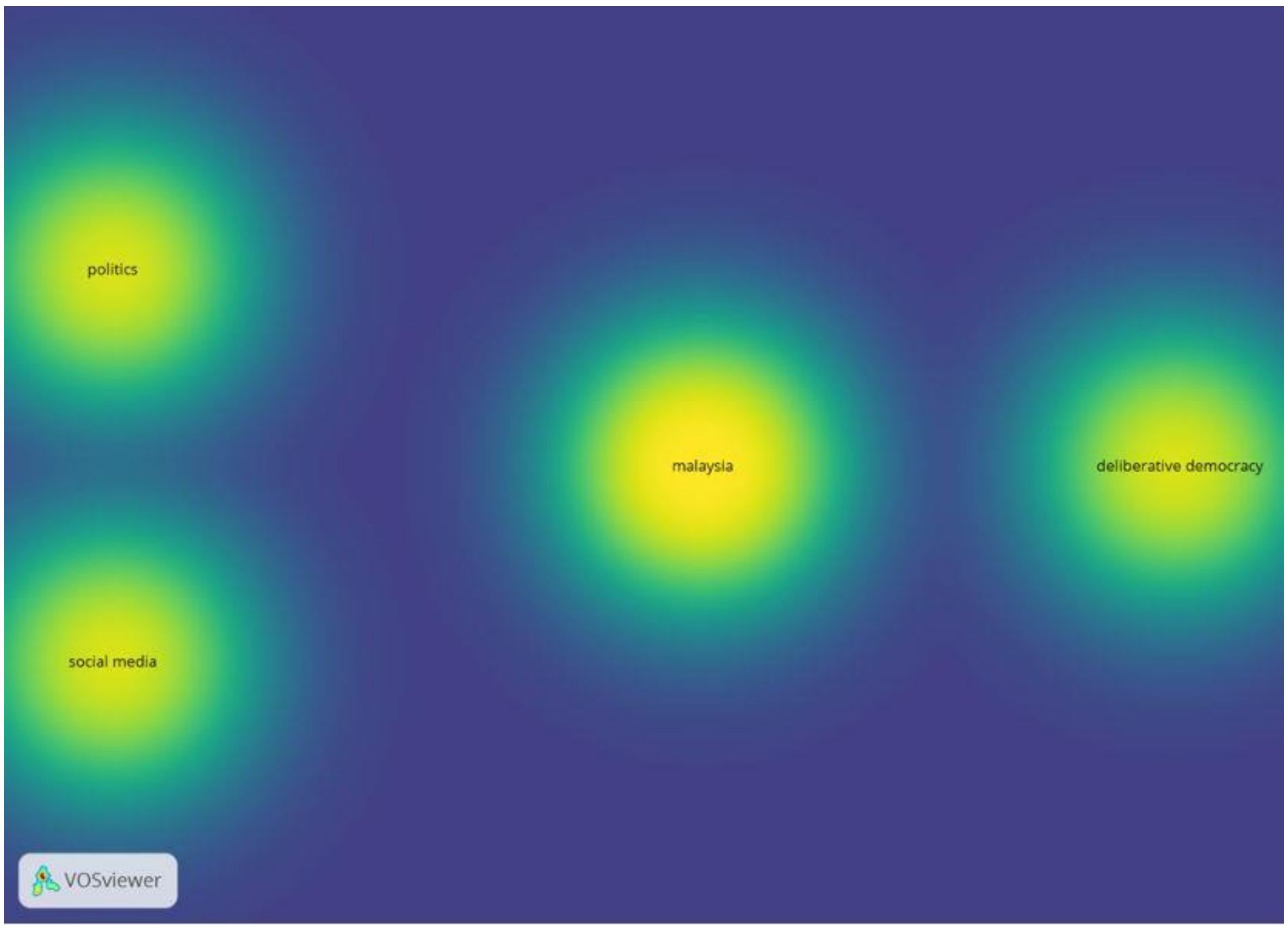

*Figure 16: Showing density network cluster of dominant themes by the top author clusters.*

## 8. Findings

The blogosphere in the Indo-Pacific region has facilitated a participatory deliberation process in various countries in the Indo-Pacific region. This is because traditional onlookers in public debates around the election, democracy, and opinion framing have taken advantage of the blogs in participatory democracy to demand good governance and demonstrate against bad government. However, public sphere deliberation shows in the study that people follow more blogs and public sphere conversations that reinforces their traditional view and personality. The opportunity to act as a commentator or as an online cybertrooper has led to people needing to be more open and aware of opposing views, howbeit to either win a new audience or antagonistic towards competing views. One of the findings in this work is also "shame labeling" of people of opposing views and silencing divergent views of minority opinion within a nucleus of a deliberation.

Another important finding of this work is the concept of cybertrooper commentators who usually flood any thread opposing their view with hate speech and anti-deliberate actions. Also, This work found that public onslaughts are traditionally one of the mechanisms opposing the position that cybertrooper bloggers and commentators form a collective judgment on opposing views. Public deliberation in the blogosphere also amplifies the natural divide between the various community in the Malaysia blogosphere, Philip- pines Blogosphere, and other blogospheres where there are natural, cultural differences and ideological differences, unlike the Australian and American blogospheres where differences are amplified in online deliberation is more focused on ideological differences as opposed to cultural differences in deliberations around elections, democracy or political conversation.

In America, this research agrees with the works of Pole (2006), where politicians participate in congressional blogging as congress members use blogs to communicate for various reasons like position-taking and advertising their works. Also, we categorize congressional bloggers as self-referenced or self-recruited cybertrooper and self-recruited commentator bloggers. Since the congress members use this means the same way that online recruits of bloggers in countries like Malaysia and the Philippines work but perform this by themselves.

## 9. Limitations

The limitation of this work comes from the fact that the volume of work done in the study area has divergent themes, which are not easy to capture in a single survey and need a consistent indexing approach returned by the search engineer. This work is aware that it may have captured only some of the various sub-themes under the Indo-Pacific blogosphere. However, the work since focuses on the various aspect and tricks in which communication and deliberation within the Indo-Pacific region public sphere are shaped by blogs and other activities by online citizens to make democracy participatory. Also, we want to give research background information that will help direct their future studies.

## 10.Conclusion & Future Work

The foundation for public deliberation, e-democracy, and online cyber advocacy of democracy and good governance was birthed by the widespread adoption of blogs and new media as a means of requesting good governance by the citizens who are traditionally not given an audience by the traditional media. Various researchers have explored concepts that have shaped these demands for good governance and the uncensored voice of the citizens. We extrapolate the historical journey of how new media, specifically blogs and people participating in the blogosphere, shaped various aspects of public deliberation.

Furthermore, we highlight the various types of bloggers engaged by various autocratic governments and opposition parties, informing cybertroopers and commentator bloggers who are more educated citizens interested in various events happening in their society. We also established the pattern of the cybertrooper group to form narratives and counter-narratives. These groups are usually financed by the ruling party or third-party public relations companies employed to help any individual push an online campaign. Existing research seems to focus on the cybertroopers as an extension of the incumbent government, but this has an exception since opposition parties with similar interests can also finance various bloggers to act as their cybertroopers. Existing research also portrays commentator bloggers as citizens who demand good governance. This is true that many commentator bloggers are enlightened citizens who have become online activists by commenting on various blog threads. These bloggers could also be fans or recruits of various political views of the incumbent governments who can engage in dialogue without acting like an online troop.

Future works should leverage how bots and cybertroopers can be used concurrently to diffuse meaningful public sphere deliberation by various political actors. Also, future work should study the link between the offline public sphere as migrated from an online community by asking important questions like how political actors take the conversation from online to offline to either sustain a political view, keeps a political party in power by winning an election and how opposition party can take advantage of the blogosphere by pushing information beliefs from online to offline. These should also include the study of how hyperlinks play roles in public sphere deliberation and how hyperlinks are used to grow the audience of blogs that acts as a secondary recruitment ground for the radicalization of online citizens. Congress members and politicians say that blogs are a dominant feature of the American blogospheres regarding public deliberation, advertising their work, and running publicity campaigns.

## ACKNOWLEDGMENT

This research is funded in part by the U.S. National Science Foundation (OIA-1946391, OIA-1920920, IIS-1636933, ACI-1429160, and IIS-1110868), U.S. Office of the Under Secretary of Defense for Research and Engineering (FA9550-22-1-0332), U.S. Office of Naval Research (N00014-10-1-0091, N00014-14-1-0489, N00014-15-P-1187, N00014-16-1-2016, N00014-16-1-2412, N00014-17-1-2675, N00014-17-1-2605, N68335-19-C-0359, N00014-19-1-2336, N68335-20-C-0540, N00014-21-1-2121, N00014-21-1-2765, N00014-22-1-2318), U.S. Air Force Research Laboratory, U.S. Army Research Office (W911NF-20-1-0262, W911NF-16-1-0189, W911NF-23-1-0011), U.S. Defense Advanced Research Projects Agency (W31P4Q-17-C-0059), Arkansas Research Alliance, the Jerry L. Maulden/Entergy Endowment at the University of Arkansas at Little Rock, and the Australian Department of Defense Strategic Policy Grants Program (SPGP) (award number: 2020-106-094). Any opinions, findings, and conclusions or recommendations expressed in this material are those of the authors and do not necessarily reflect the views of the funding organizations. The researchers gratefully acknowledge the support.